\documentclass[aps,12pt,tightenlines]{revtex4}
\usepackage[dvips]{color,graphicx}
\begin{document}
\newcommand{\boldsigma}{\mbox{\boldmath $\sigma$}}
\newcommand {\boldgamma}{\mbox{\boldmath$\gamma$}}
\newcommand{\boldtau}{\mbox{\boldmath $\tau$}}
\newcommand{\bftau}{\mbox{\boldmath $\tau$}}
\newcommand{\bfx}{{\bf x}}
\newcommand{\bfy}{{\bf y}}
\newcommand{\bfP}{{\bf P}}
\def\poinc{Poincar\'{e} }
\newcommand{\eq}[1]{Eq.~(\ref{#1})}
\def\bfq {{\bf q}}
\def\bfm {{\bf m}}
\def\bfs {{\bf s}}
\def\bfn {{\bf n}}
\def\bfqp {{\bf q}_\perp}
\def\bfK{{\bf K}}
\def\bfKp{{\bf K}_\perp}
\def\bfL{{\bf L}}
\def\bfk{{\bf k}}
\def\bfp{{\bf p}}  
\newcommand{\bfkap}{\mbox{\boldmath $\kappa$}} 
\def\bfr{{\bf r}} 
\def\bfy{{\bf y}} 
\def\bfx{{\bf x}} 
\def\be{\begin{equation}}
 \def \ee{\end{equation}}
\def\bea{\begin{eqnarray}}
  \def\eea{\end{eqnarray}}
\newcommand{\eqn} {Eq.~(\ref )}
\newcommand{\bb}{\langle}
\newcommand{\kk}{\rangle}
\newcommand{\bk}[4]{\bb #1\,#2 \!\mid\! #3\,#4 \kk}
\newcommand{\kb}[4]{\mid\!#1\,#2 \!\mid}

\def\notp{{\not\! p}}
\def\notk{{\not\! k}}
\def\up{{\uparrow}}
\def\down{{\downarrow}}
\def\bfb{{\bf b}}

\setlength{\textheight}{8.60in}
\setlength{\textwidth}{6.6in}
\setlength{\topmargin}{-.40in}
\setlength{\oddsidemargin}{-.125in}
\tolerance=1000
\baselineskip=14pt plus 1pt minus 1pt

\def\poinc{Poincar\'{e} }
\def\bfq {{\bf q}}
\def\bfK{{\bf K}}
\def\bfL{{\bf L}}
\def\bfk{{\bf k}}
\def\bfp{{\bf p}}  
\def\be{\begin{equation}}
 \def \ee{\end{equation}}
\def\bea{\begin{eqnarray}}
  \def\eea{\end{eqnarray}}
\def\eqn {Eq.~(\ref )}

\newcommand{\kx}[2]{\mid\! #1\,#2 \kk}
\def\notp{{\not\! p}}
\def\notk{{\not\! k}}
\def\up{{\uparrow}}
\def\down{{\downarrow}}
\def\bfb{{\bf b}}

\vspace{1.0cm}
\vspace{.50cm}
\title{\begin{flushright}{\normalsize NT@UW-06-16}\end{flushright}
Detecting Strangeness $-4$ Dibaryon  States}

\author{Gerald A. Miller}
\affiliation{ University of Washington
  Seattle, WA 98195-1560}

\sloppy


\begin{abstract} Recent experiments at Jefferson Laboratory and potential new facilities  at the
Japan Proton Accelerator Research Complex (J-PARC) make it evident that 
 the discovery of a
$^1S_0$ di-cascade bound state of two $\Xi$ particles is   feasible. We state the simple
arguments, based on $SU(3)$  flavor symmetry, 
for the existence of this  bound state, review the previous predictions and comment
on the experimental conditions  necessary for detection.
\end{abstract}
\maketitle
\vskip0.5cm


The  first measurement of exclusive doubly-strange cascade  $\Xi^-$(1321) 
hyperon production in the $\gamma p\rightarrow K^+K^+\Xi^-$ 
reaction at Jefferson Laboratory has recently been reported \cite{Price:2004hr}.
The high-quality photon beam was used to excite the narrow $\Xi^-$ state that
was observed as a sharp peak in the missing mass spectrum. This success opens the
door to many avenues of research including double hypernuclear production.

The interest in understanding the properties of the cascade spectrum stem from
$QCD$ which, in its earliest incarnation, 
expressed the strong-interaction Hamiltonian as the sum of an $SU(3)$ 
invariant term and a medium strong interaction term (now
known as the quark mass matrix) that breaks the $SU(3)$ \cite{gman}. This reasoning
led to the understanding of baryon level 
spacings--the Gell-Mann-Okubo mass formula, and many other successes in
understanding strong and electromagnetic interactions \cite{eight}. The Gell-Mann-Okubo mass
formula shows that possible modifications of baryonic wave functions, caused by the difference
between the strange and light quark masses, do  not modify the spectrum.
Our purpose here is apply the   old theoretical insights to the two-baryon
system. In particular we shall argue for the likely existence of a $^1S_0$
loosely bound state of two cascade particles-- the di-cascade. Then we shall discuss the newly
feasible 
reactions
that allow this system to be detected. Finding  such a strangeness -4, baryon 2
 system would be the discovery of a new dibaryon particle. Its existence would 
 verify the flavor symmetry of the $u,d$ and massive $s$ quark 
interactions for systems of  two baryons in the
same irreducible representation of $SU(3)_F$. This would   
 provide 
insight into  how QCD works. 
 The ability to  understand
strange nuclear matter would be increased and  impetus would be given to lattice QCD studies  of
two baryon interactions\cite{Beane:2006mx}.

The first step is to realize that 
$SU(3)$ flavor  symmetry predicts the equality
of the $^1S_0$ strong nucleon-nucleon NN interaction with the  $^1S_0$  $\Xi\Xi$
strong interaction because the
$NN$ and  $\Xi\Xi$ systems are each in 
the $\{27\}$ dimensional irreducible representation of $SU(3)$ \cite{Stoks:1999bz}.
This equality, the known existence of a quasibound state in the
$^1S_0$ NN channel, and the increase of the reduced mass in the $\Xi\Xi$ channel,
makes it likely that the $^1S_0$
 $\Xi\Xi$ state is bound. 

We next discuss the equality of the  $^1S_0$ and $\Xi\Xi$
 $^1S_0$ strong interactions. It is convenient to use the formalism of 
Savage \& Wise\cite{Savage:1995kv}.
The baryon fields are introduced as a $3\times 3$ octet matrix
\begin{equation}\label{baryons}
B = \left[\matrix{\Sigma^0/\sqrt{2} + \Lambda/\sqrt{6} &
\Sigma^+ & p\cr
\Sigma^- & -\Sigma^0/\sqrt{2} + \Lambda/\sqrt{6} & n\cr
\Xi^- & \Xi^0 & -\sqrt{{2\over 3}} \Lambda \cr}\right]\ \ \ ,
\end{equation}
that transforms under chiral $SU(3)_L \times SU(3)_R$ as
$B \rightarrow UBU^\dagger.$

The properties of low energy baryon-baryon interactions can be described
by 
 terms in ${\cal L}$ with four baryon
fields (and
no derivatives).  They are given by
\begin{eqnarray}\label{fourb}
{\cal L}^{(2)} =
&-&{c_1} Tr (B_i^\dagger B_i B_j^\dagger B_j)
- {c_2} Tr (B_i^\dagger B_j B_j^\dagger B_i)\nonumber\\
&-& {c_3} Tr (B_i^\dagger B_j^\dagger B_i B_j)
- {c_4} Tr (B_i^\dagger B_j^\dagger B_j B_i)\nonumber\\
&-& {c_5} Tr (B_i^\dagger B_i) Tr (B_j^\dagger B_j)
- {c_6} Tr (B_i^\dagger B_j) Tr (B_j^\dagger B_i) \ \ \
{},
\end{eqnarray}
where the indices $i,j$ represent the spin of the two-component baryon fields, and
repeated indices are summed over.
In writing \eq{fourb} we have ignored the explicit effects of the exchange of the 
pseudo-Goldstone bosons, terms of higher order in the chiral expansion and $SU(3)_F$ breaking terms.
 We shall return to all of those below.

The key feature of our present interest is that the nucleon and cascade doublets
occupy analogous positions in the 
baryon matrix \eq{baryons}. Therefore 
 the interaction \eq{fourb} is invariant under the  transformation 
$NN\leftrightarrow \Xi\Xi$. 
Evaluation of \eq{fourb}, using the properties of the Majorana
exchange operator $B_jB_i={1\over2}(1+\boldsigma_i\cdot\boldsigma_j)B_iB_j$,
and keeping only the cascade and nucleon states 
leads to the result:
\bea \label{XiN}
{\cal L}^{(2)}_{N,\Xi} =
&-&\left({c_1+c_5+{1\over2}(c_2+c_6)}\right) \left((\Xi^\dagger \Xi)^2+(N^\dagger N)^2)\right) \nonumber\\
&-& {(c_2+c_6)} {1\over2} \left(\Xi^\dagger\boldsigma\Xi\cdot \Xi^\dagger\boldsigma \Xi)+
N^\dagger \boldsigma N\cdot N^\dagger \boldsigma N)\right) \nonumber\\
&-& 2{(c_3+{1\over2}c_4)}\Xi^\dagger N^\dagger N\Xi+c_4\Xi^\dagger \boldsigma N\cdot N^\dagger\boldsigma\Xi
.
\end{eqnarray}
Equation (\ref{XiN}) makes clear the prediction of the equality of the 
$\Xi\Xi$ and $NN$ interactions, already present in \eq{fourb}.  The interactions
of \eq{XiN} refer to both the $^1S_0$ and $^3S_1$ channels. However the
one pion exchange interaction has a big influence in the triplet channel, and
a much smaller influence in the singlet channel. Indeed,
 KSW\cite{Kaplan:1998tg}
 counting (in which the pion exchange interaction is treated as a perturbation)
may be applied to the $^1S_0$ but not the $^3S_1$ channel \cite{Beane:2001bc}.
Furthermore the $NN$ and $\Xi\Xi$ $^3S_1$ states do not belong to the same irreducible 
representation of $SU(3)$ \cite{Stoks:1999bz}.
Therefore we shall consider only  the $^1S_0$ channel, and
neglect 
the  explicit  effects of 
one boson exchange as well as the possibly important effects of flavor symmetry breaking 
of interactions in the
schematic calculations we present here. Both of these effects
  are taken into account in 
realistic calculations \cite{Stoks:1999bz} that have obtained  
the same conclusions that 
we shall reach below.

The Lagrangian \eq{XiN} must be extended by including the kinetic energy term 
 ${\cal L}_{KE} =- Tr B_i^\dagger(\nabla^2/2M_B)
B_i$. In addition there are four-baryon terms that 
involve derivative operators. We shall not include  such terms (the $d_i$ terms)
 directly.
Our purpose here is to obtain simple interactions
and then show that the use of different reduced masses in the Schroedinger equation
can lead to bound states. Therefore we study the low energy regime in which
the interaction is well described by the scattering length $a$ and effective
range $r_e$ so that the phase shift $\delta(k)$ can be expressed as 
\bea k \cot \delta=-{1\over a} +{1\over2}r_ek^2.\eea 
In this regime including the effects of the $d_i$ terms is indistinguishable from 
using a  nucleon-nucleon
 simple potential that is defined by a depth and a range. Here we shall 
consider  three potentials:
 square well of depth $V_0$ and range  $R$, 
a non-local separable potential of the form 
$V(r,r')= -{\lambda\over 2\mu} u(r)u(r')$, where $u(r)={e^{-r\over b}\over r},$ 
$\mu$ 
is the $NN$ effective mass,  and
a delta-shell potential $V(r)=-\lambda \delta(r-R)$. The latter two are taken
from the text by Gottfried\cite{gott}. One may choose the depth and range parameter
to reproduce the scattering length and range of the $^1S_0$ system. We use
either $a=-18$ fm (from 
the average of the $nn$ and $pp$ systems, set I) or 
$a=-24$ fm from the $np$ system (set II). In either case we take 
$r_e= 2.8$ fm.  The large magnitude and negative sign of the scattering length
 indicates the presence of a quasibound state: a slightly stronger interaction
between nucleons would have caused a bound state 
to appear and the scattering length to be positive. The Schroedinger equation
in the $^1S_0$ channel can be expressed (for a local potential) as
\bea -{d^2u\over dr^2} + 2\mu V u =k^2 u,\eea
where $u(r)/r$ is the wave function. If the $\Xi\Xi$ interaction is the
same as the nucleon-nucleon interaction $V$, then using the appropriate $\Xi\Xi$
reduced mass corresponds to a forty percent increase in the strength of 
the interaction. Alternately, the square of the effective momentum  inside the
well $k^2-2\mu V$ would  be increased.

Obtaining analytic expressions for the scattering length and effective range
is a straightforward matter for each of the potentials we employ. The 
results for the square well $sq$  are:
\bea
a_{sq} = R {x \cot x-1\over x \cot x},\; r_e^{sq}= {3 x +(3-6x^2)\cot x +
x(-3+2x^2)\cot^2x\over
3 x(-1+x\cot x)^2}, x\equiv R\sqrt{2\mu V_0} \eea
To achieve a negative scattering length of very large magnitude the 
value of $x$ must be slightly less than 
${\pi\over2}.$ If the value of $x$ were to be slightly greater 
than ${\pi\over2},$ then $a$ would be positive and a bound state would
exist. We find that
$x=1.48, R=2.64$ fm to reproduce set I, or $x=1.50, R=2.68$ fm, to reproduce
set II.  If $SU(3)$ flavor symmetry holds for the interaction $V$ the values of
of $x_{\Xi\Xi}$ for the cascade system would be either  1.76 (set I) or 1.78 
(set II).
These correspond to scattering lengths of 10.6 fm, and 9.81 fm, and 
binding energies of 7.48 MeV and 6.83 MeV.

The analysis of the separable potential $sep$ case proceeds in a similar manner
with similar results. The results for the phase shift are in  ref.~\cite{gott}:
\bea a_{sep}={2\xi b\over \xi-1},\;r_e^{sep}={b (2+\xi)\over \xi },\;\xi=2\pi\lambda b^3.\eea
To achieve a negative scattering length of very large magnitude the 
value of $\xi$ must be slightly smaller than one. If the value of $\xi$ were to
be slightly greater than one, then $a$ would be positive and a bound state would
exist.   We find that
$\xi=0.911, b=0.88$ fm to reproduce set I, or $\xi=0.931, b=0.89$ fm, to reproduce
set II.  If $SU(3)$ flavor symmetry holds for the interaction $V$ the values of
of $\xi_{\Xi\Xi}$ for the cascade 
system would be either  1.28 (set I) or 1.30 (set II).
These correspond to scattering lengths of 8.0 fm, and 7.6 fm.
The binding energy $B={\alpha^2\over 2\mu_{\Xi\Xi}}$ is obtained by solving the
equation \bea
1=\xi_{\Xi\Xi}{1-2 y +y^2\over (y^2-1)^2},\; y\equiv \alpha b.\eea
We find $\alpha b=0.131$ and $0.140$  that correspond to 
binding energies of 0.66 MeV and 0.73 MeV.

The analysis of the delta-shell potential $dsh$ case is also similar.
 The  phase shifts  are  presented in  
Ref.~\cite{gott}, with  
\bea a_{dsh}=R{\gamma\over \gamma-1},\;r_e^{dsh}={{2\over3}R (1+\gamma)\over \gamma },\;\gamma\equiv
\lambda R.\eea
To achieve a negative scattering length of very large magnitude the 
value of $\gamma$ must be slightly smaller than one, with a  value 
 slightly greater than one corresponding to the existence of  a bound state.
   We find that
$\gamma=0.930, R=2.02$ fm to reproduce set I, or $\gamma=0.946, R=2.04$ 
fm, to reproduce
set II.  If $SU(3)$ flavor symmetry holds for the interaction $V$ the values of
of $\gamma_{\Xi\Xi}$ for the cascade 
system would be either  1.30 (set I) or 1.32 (set II).
These correspond to scattering lengths of 8.75 fm, and 8.34 fm.
The binding energy $B={\alpha^2\over 2\mu_{\Xi\Xi}}$ is obtained by solving the
equation \bea
\gamma_{\Xi\Xi}= \alpha R (1+\coth \alpha R).\eea
We find $\alpha R=0.275$ and $0.291$  that give  
binding energies  of 0.549 and 0.606 MeV.

For each of the models considered, the use of a given potential combined with changing the
reduced mass from that of $NN$ to that of $\Xi\Xi$ leads to the prediction of a positive 
scattering length and the existence of a bound state. The scattering lengths range from about eight to eleven
fm, while the binding energies have a much wider range from about 0.55 to 7.5 MeV. These calculations use
very simple potentials, but the argument is clear. Increasing the magnitude of a  potential that just misses 
having a bound state by forty percent should lead to the existence of a bound state.

It is necessary to discuss the effects of including the pseudo-Goldstone bosons that appear in 
chiral perturbation theory
as well as $SU(3)$ breaking terms that enter in the interactions. These effects are included 
in Ref.~\cite{Stoks:1999bz}
which obtained
  soft-core baryon baryon potentials for the complete baryon octet using the formalism 
of Ref.~\cite{Rijken:1998yy}.
The potentials are parameterized in terms of one-boson exchanges. Boson-nucleon form factors 
are included to
handle the short-distance part of the interaction. 
The form factors depend on the $SU(3)_F$ assignment of the mesons. The
$^3P_0$ mechanism is used to generate the flavor-symmetry breaking of the coupling constants. Six different models of the hyperon-nucleon interaction that 
describe the data equally well
 are constructed in Ref.~\cite{Rijken:1998yy}. All of the parameters of each model are fixed in  Ref.~\cite{Rijken:1998yy}
so that each defines a baryon-baryon model that models all possible baryon-baryon 
interactions \cite{Stoks:1999bz}.
Each of the six potentials predicts the existence of a $\Xi\Xi$ bound state in the $^1S_0$ channel. 
The binding energies range
from 0.1 to 15.8 MeV, a variation that is similar to that obtained using simple potentials.

Note also that the existence of $\Lambda\Lambda$ hypernuclei, taken along
with $SU(3)_F$ symmetry, implies that the hyperon-hyperon interaction is strongly
attractive \cite{Schaffner:1993qj}. One may construct one boson exchange potentials
that reproduce the strong attraction in the $\Xi\Xi$ channel, 
providing another model\cite{Schaffner:1993qj}.

We have seen that a  $\Xi\Xi$ $^1S_0$ bound state (di-cascade) 
 occurs in at least 
six  realistic and three simple (a total of nine) different potential models. Its existence
 is therefore more than plausible,
 so we next comment briefly about  properties and methods of detection. 
These loosely-bound di-cascade states  would decay by the weak interaction to $NN4\pi$ final states. The
 lifetime would be roughly that of a free $\Xi$, about $2 \times 10^{-10}$ s.
For a discussion of other weak decay modes see Ref.~\cite{Schaffner-Bielich:1999sy}. Furthermore, the
small binding energies tell us that the di-cascade  
consists of two well-separated baryons and therefore is  fragile and
easily absorbed if  produced in a reaction that surrounds it with  nucleons.

We
concentrate on the $\Xi^0\Xi^-$ or $\Xi^0\Xi^0$ systems because the repulsive effects
of the Coulomb interaction could cause a state, weakly bound 
 under the strong interaction, to be unbound.  Reactions involving two baryons seem
best suited for the discovery of the $^1S_0$ bound state. Therefore it seems promising to make the
search   at Jefferson Laboratory using the missing-mass
technique in the reactions
\bea \gamma +D\rightarrow (\Xi^0\Xi^-)_{^1S_0}+ K^++K^++K^0K^0\\
\gamma +D\rightarrow (\Xi^0\Xi^0)_{^1S_0}+ K^++K^0+K^0K^0.
\eea
The photon threshold energy is about 4.8 GeV. This reaction presents 
the difficulty of measuring four kaons, but
there should be a clear signature as a sharp peak in the missing-mass spectrum.
The future availability of high intensity $K^-$ 
 beams at J-PARC makes it interesting to consider the reactions
\bea K^- +D\rightarrow (\Xi^0\Xi^-)_{^1S_0}+ +K^++K^0K^0\\
K^- +D\rightarrow (\Xi^0\Xi^0)_{^1S_0}+ +K^0+K^0K^0.
\eea
The kaon threshold energy is 3.8 GeV, and one would need to  detect only three kaons in the final state.

Both the photon and kaon induced searches would require 
much experimental effort to find the $(\Xi\Xi)_{^1S_0}$ di-cascade bound state.
 However, the recent observation of the $\Xi^-$ in the $\gamma p$ reaction at Jefferson Laboratory
and the expected availability of high-intenstiy kaon beams at J-PARC make it evident that  the necessary 
experimental tools  exist  or can be obtained. This  bound state is predicted to exist in nine different
models, so that a careful search is likely to be successful. However, 
 models can not provide a definitive proof that the state of interest exists. 
Therefore we call for the development of experiments capable of determining 
 whether or not the  $(\Xi\Xi)_{^1S_0}$ di-cascade bound
state exists.

\section*{Acknowledgments}
I thank the USDOE  
  for partial support of this work. I thank Martin Savage and Ben Nefkens for 
useful discussions.

\end{document}